# Charge localization at the interface between $La_{1-x}Sr_xMnO_3$ and the "infinite layers" cuprate $CaCuO_2$

Nan Yang,[1] D. Di Castro,[1,a)] C. Aruta,[2] C. Mazzoli,[3] M. Minola,[3] N. Brookes,[4] M. Moretti Sala,[4] W. Prellier,[5] O. I. Lebedev,[5] A. Tebano,[1] and G. Balestrino[1]

[1]*CNR-SPIN and Dipartimento di Ingegneria Informatica Sistemi e Produzione,Università di Roma Tor Vergata, Via del Politecnico 1, I-00133 Roma, Italy*

[2]*CNR-SPIN, Dipartimento di Scienze Fisiche, Via Cintia, Monte S.Angelo, 80126 Napoli, Italy*

[3]*CNISM and Dipartimento di Fisica, Politecnico di Milano, I-20133, Italy*

[4]*European Synchrotron Radiation Facility, 6 rue Jules Horowitz, BP 220, 38043 Grenoble, Cedex 9, France*

[5]*Laboratoire CRISMAT, UMR 6508, CNRS-ENSICAEN 6Bd Marechal Juin, 14050 Caen, France*

$(CaCuO_2)_m/(La_{0.7}Sr_{0.3}MnO_3)_n$ superlattices, consisting of the infinite layers cuprate $CaCuO_2$ and the optimally doped manganite $La_{1-x}Sr_xMnO_3$, were grown by pulsed laser deposition. The transport properties are dominated by the manganite block. X-Ray Absorption spectroscopy measurements show a clear evidence of an orbital reconstruction at the interface, ascribed to the hybridization between the Cu $3d_{3z^2-r^2}$ and the Mn $3d_{3z^2-r^2}$ orbitals via interface apical oxygen ions. Such a mechanism localizes holes at the interfaces, thus preventing charge transfer to the $CaCuO_2$ block. Some charge (holes) transfer occurs toward the $La_{0.7}Sr_{0.3}MnO_3$ block in strongly oxidized superlattices, contributing to the suppression of the magnetotransport properties.

---

a) Electronic mail: daniele.di.castro@uniroma2.it

Heteroepitaxial structures based on strongly electron correlated oxides are attracting an increasing attention because of their possible practical applications in the emerging field of oxide electronics.[1,2] Namely, the interaction at the interface between the constituent oxides can result in a number of exotic properties, including the occurrence of two-dimensional phases of electron matter at the interface between insulating oxides,[3 , 4] low transition temperature ($T_c$) superconductivity,[5] high $T_c$ superconductivity.[6,7,8] A wealth of microscopic mechanisms may be at the work at the interface between different oxides. These include discontinuity in the polar sequence of the atomic layers at the interface,[9] orbital reconstructions induced by misfit strain and/or reduced dimensionality,[10 , 11] exchange interactions across the interface, and electrical charge transfer.

Since cuprates and manganites have similar in plane lattice spacing, their combination in a single heterostructure has been especially studied.[12,13,14,15,16] Based on a detailed analysis of available photoemission and diffusion voltage experiments, doping of high-$T_c$ cuprates by charge transfer from manganites or other oxides using heterostructure architectures has been investigated theoretically in ref. 17. The major conclusion was that addition of carriers to antiferromagnetic Cu oxides may lead to a superconducting state at the interface.

On the other hand the experimental scenario about charge transfer in cuprates/manganites heterostructures is still controversial. A depression of the critical temperature $T_c$ was experimentally observed in $YBa_2Cu_3O_{7-x}/La_{0.67}Ca_{0.33}MnO_3$ superlattices when the $YBa_2Cu_3O_{7-x}$ (YBCO) thickness is reduced.[18,19] Tentatively, the $T_c$ decrease was attributed to the role of the ferromagnetic exchange field in reducing the pairs formation in the YBCO layers. A strong antiferromagnetic Cu-Mn exchange coupling was also observed. More recently, the local orbital structure at the interface between YBCO and manganite was investigated by spatially and element resolved spectroscopic techniques. [11] It was found that the electronic structure of the $CuO_2$ planes is modified by covalent bonds between Cu and Mn ions at the interface.

In our opinion, one of the major difficulties to overcome in order to fully exploit the possibilities offered by cuprate/manganite heterostructures, is the structural complexity of the cuprate block. In most practical cases it consists of the YBCO structure, where two non equivalent Cu-O planes, two different spacers between the Cu-O planes (Y and BaO planes) and a build-in charge reservoir (the $CuO_x$ layer between two BaO planes) are present. Such a degree of structural complexity, together with the multiplicity of possible interactions at the interface, make difficult the unambiguous interpretation of the experimental results.

We thus propose a different and simplified approach to the problem based on the use of a cuprate with "infinite layers" (IL) structure for the engineering of cuprate/manganite heterostructures. The IL $CaCuO_2$ (CCO) is the simplest antiferromagnetic parent compound of cuprate superconductors. [20] It has the big advantage of consisting exclusively of $CuO_2$ planes separated by bare Ca atoms, thus avoiding any complication related to non equivalent Cu sites or build-in charge reservoirs. It was shown that, if grown at very high pressure, the Sr doped CCO pellet could become a high temperature superconductor, because of the occurrence of SrO planar defects substituting $CuO_2$ planes. [21] Very recently, pure CCO has been used to synthesize $CaCuO_2/SrTiO_3$ heterostructure films. [8] If grown in strongly oxidizing conditions, these haterostructures become superconducting at about 40K, since extra oxygen atoms enter at the interface, act as apical oxygen for the Cu in the $CuO_2$ planes of CCO, and provide holes for doping. This sort of interface reconstruction thus allows a charge transfer from the interface layers to the CCO block. In this work we used CCO in combination with the manganite $La_{0.7}Sr_{0.3}MnO_3$ (LSMO) to synthesize $(CaCuO_2)_m/(La_{0.7}Sr_{0.3}MnO_3)_n$ superlattices (SLs), where $m$ and $n$ are the number of unit cells of $CaCuO_2$ and $La_{0.7}Sr_{0.3}MnO_3$, respectively. We found that these systems have a high quality superlattice structure and can be a very good candidate for investigating interface reconstruction phenomena and the possible occurrence of charge transfer at the cuprate/manganite interface.

We synthesized several $La_{0.7}Sr_{0.3}MnO_3$ films and $(CaCuO_2)_m/(La_{0.7}Sr_{0.3}MnO_3)_n$ superlattices (where $m$ and $n$ are the number of unit cells of $CaCuO_2$ and $La_{0.7}Sr_{0.3}MnO_3$, respectively) by pulsed laser deposition (KrF excimer laser, $\lambda = 248$ nm) on $NdGaO_3$ (110) (NGO) oriented mono-crystalline substrates, NdO surface terminated.[22] Two targets, with $CaCuO_2$ and $La_{0.7}Sr_{0.3}MnO_3$ nominal composition, mounted on a multitarget system, were used. Substrate temperature during the deposition was T $\approx 600°C$. Films and SLs were deposited in two different oxidizing conditions: i) $8 \times 10^{-1}$ mbar of $O_2$ followed by quenching to room temperature in 1 bar of $O_2$ (moderately oxidizing conditions) and ii) $8 \times 10^{-1}$ mbar of $O_2$/12% $O_3$ followed by quenching to room temperature in 1 bar of $O_2$ (strongly oxidizing conditions). The latter procedure was aimed at increasing as much as possible the oxygen doping of the SLs. Structural characterization was performed by Transmission Electron Microscopy (TEM) and x-ray diffraction (XRD) in $\theta$-$2\theta$ Bragg-Brentano geometry. TEM analysis was performed by using a Tecnai G2 30 UT microscope operated at 300 kV and having 0.17 nm point resolution. Fig.1 shows cross-section low magnification TEM image (a), corresponding electron diffraction (ED) pattern (Fig.1b), and high resolution TEM (HRTEM) image (Fig.1c) for a $(CaCuO_2)_3/(La_{0.7}Sr_{0.3}MnO_3)_{16}$ superlattice. Low magnification TEM and corresponding ED pattern clearly show heteroepitaxial growth of the superlattice. A typical series of stacked layers with different contrast (bright corresponds to CCO, dark corresponds to LSMO layer), having regular alternative thicknesses, are observed. HRTEM image (Fig. 1c) confirms an heteroepitaxial growth of CCO and LSMO layers and shows sharp interfaces. The superlattice period $\Lambda$ was estimated from intensity scan profile (Fig.1d) of HRTEM image and is about 79 Å. The XRD spectra, around the substrate (002) reflection, for selected SLs $(CaCuO_2)_3/(La_{0.7}Sr_{0.3}MnO_3)_n$ with $n$ ranging from 4 to 16, are shown in Fig. 2. The good quality of the superlattice structure is clearly confirmed by the presence of sharp high order satellite peaks $SL_{\pm i}$ around the average structure peak $SL_0$. All films were found to have a mosaic spread ($\approx 0.07°$) coinciding with the substrate value. From the position of the satellite peaks $SL_{-1}$ and $SL_{+1}$, it is possible to evaluate accurately the thickness of the supercell $\Lambda = m \cdot c_{CCO} + n \cdot c_{LSMO}$, where $c_{CCO}$ is the c-axis parameter of CCO and $c_{LSMO}$ the one of LSMO. With increasing the number of LSMO unit cell, both $SL_0$ and $SL_{\pm i}$ evolve in the appropriate way ($2\theta$ for $SL_0$ decreases and the distances between satellite peaks $SL_{-i}$ and $SL_{+i}$ shrink). The individual layer thicknesses $n$ and $m$ were then calculated following the approach outlined in ref. 8. The experimental error on the layer thickness was conservatively estimated to be one unit cell. From the XRD spectrum of the same sample used for HRTEM analysis, the values $n = 16 \pm 1$ u.c. and $\Lambda = 75 \pm 2$ Å have been found, in very good agreement with the HRTEM results.

For all samples, regardless of the thickness $m$ and $n$ of the individual blocks, the electrical transport properties are dominated by the LSMO block. Namely, $\rho$ versus T ($\rho(T)$), measured in the Van der Pauw geometry, for the $CCO_m/LSMO_n$ SLs shows roughly the same behavior as for bare LSMO films $n$ u.c. thick. The metal-insulator transition temperature ($T_p$) is well above room temperature for $n>18$ u.c. and then gradually decreases as the thickness is reduced, finally disappearing below a critical value ("dead layer" effect). Such a critical thickness was estimated in 6 u.c. for the bare LSMO films.[10] The origin of such a phenomenon in ultrathin LSMO films has been investigated[10,20-22] by Linear Dichroism (LD) of X-Ray Absorption Spectroscopy (XAS) in few unit cells thick LSMO films on NGO. An orbital reconstruction with preferential *3d $e_g$ (3$z^2$-$r^2$)* Mn orbital occupation was found,[10,23] ultimately stabilizing the antiferromagnetic C-type phase.[24,25] In Fig. 3 $\rho$(T) is reported for two $CCO_3$/$LSMO_{14}$ SLs, grown under strongly oxidizing conditions (SOC) and moderately oxidizing conditions (MOC) (curves *a)* and *c)*, respectively) and for two LSMO films, about 15 unit cells thick, grown under SOC and MOC (curves *b)* and *d)*, respectively). Furthermore, the $\rho$(T) behavior is reported also for a $CCO_9$/$LSMO_4$ SL (curve *e)*) grown under SOC. The behavior of $\rho$(T) is similar for *b)*, *c)* and *d)*, while $\rho$(T) for *a)* is clearly different: the temperature of the metal insulator transition, $T_p$, is 280 K (to be compared with 340 K for *b)*, *c)* and *d)*). Furthermore, the low temperature resistivity of *a)* is definitely larger. Curve *e)* shows an insulating behavior over the whole temperature range, as expected for a LSMO layer of 4 u.c. This latter finding gives further evidence that the CCO layer, regardless of its thickness, does not play a relevant role in the conduction of CCO/LSMO SLs.

In Fig. 4, $T_P$ as a function of the thickness (in u. c.) of the LSMO block is reported for several samples. The full line represents the behavior for bare LSMO films. To draw this curve, experimental data from the present work (full triangles) and from ref. 10 have been used. For bare LSMO films no dependence on the oxidizing conditions (SOC or MOC) was detected. The full black squares and empty triangles represent CCO/LSMO SLs grown under SOC and MOC, respectively. While the empty triangles, within the experimental error, do not depart from the full line, the full black squares definitely show a different behaviour: namely, the suppression of the magnetotransport properties in the case of strongly oxidized SLs takes place for larger layer thicknesses.

To understand the mechanism at the basis of the behaviour described above we have carried out XAS and LD XAS measurements. LD is defined as the difference V-H between the XAS spectra taken in vertical (V) polarization (electric field vector **E** parallel to the sample surface) and in horizontal (H) polarization (electric field vector **E** forms a 30° angle with the surface normal). As a consequence, a positive LD is roughly an indication of in-plane holes excess. XAS and LD XAS measurements were carried out at both the Cu $L_3$ and the Mn $L_{2,3}$ edge at the ID08 beam line of European Synchrotron Radiation Facility. The measurements were recorded by collecting the sample drain current in total electron yield (probing depth about 10nm) and performed with the x-ray beam at 30° from the sample surface.

XAS is an element sensitive technique that allows investigating separately the CCO and the LSMO block. Figs. 5 *a)* and *b)* show the normalized XAS spectra taken at the Cu $L_3$-edge (Cu 2p --> Cu 3d electron transition) for a CCO$_3$/LSMO$_{14}$ superlattice and a reference CCO film, respectively. In the case of the CCO film the $L_3$ peak at 932.5 eV is assigned to the absorption by an undoped Cu site. In hole doped cuprates[26,27] and cuprate based superlattices[8,28] a shoulder, attributed to the $3d^9\underline{L}{\rightarrow}3d^{10}\underline{L}$ transition, where $\underline{L}$ indicates an additional oxygen ligand hole arising from Cu 3d - O 2p hybridization, develops about 1.5 eV above the main peak. No trace of such a shoulder is seen in the spectrum, in agreement with the circumstance that CCO is an undoped insulating cuprate. The XAS signal from CCO is strongly anisotropic: the absorption in H-polarization (**E**//c) is much weaker than that in V-polarization (**E**//ab). Such behaviour is typical of the "infinite layers" structure and is connected to the purely planar oxygen coordination of the Cu ions in this structure that populates with holes the $3d_{x2-y2}$ orbital at the expenses of the $3d_{3z2-r2}$ orbital. In the Cu absorption spectrum of the CCO$_3$/LSMO$_{14}$ superlattice grown under SOC, no $3d^9\underline{L}{\rightarrow}3d^{10}\underline{L}$ shoulder develops on the high energy side of the main peak, demonstrating that the CCO block in the SL remains undoped. On the contrary, with respect to the CCO film, a drastic decrease of the LD signal and a shift of the main peak to lower energy values occur. A similar behaviour was observed at the interfaces in different cuprate/manganite superlattices (YBCO/LCMO and PrYBCO/LCMO) and attributed to covalent bonding at the interface between *Cu* and *Mn $3d_{3z^2-r^2}$* orbitals via apical oxygen ions.[11] Namely, covalent bonding gives rise to an orbital reconstruction that results in an increase of the *Cu $3d_{3z^2-r^2}$* holes population and, therefore, in a sizeable decrease of the dichroic signal.

The XAS spectrum at the *$L_{2,3}$-Mn* edge is reported in Fig.6 a) for two different CCO/LSMO strongly oxidized SLs having the same CCO block (3 u.c.) but different LSMO thickness, namely 4 and 14 u.c.. For the former sample the signal comes mostly from the interface, while for the latter there is a sizeable contribution from the bulk of LSMO block. The major difference between the spectra consists in the appearance of a small peak in the more interface sensitive spectrum of CCO$_3$/LSMO$_4$ (indicated by an arrow in Fig. 6), which can be attributed to excess *$Mn^{4+}$* (Ref. 29). For both superlattices, a sizeable dichroic signal is detected (Fig. 6 *b)*), with a clear signature of the *$3z^2$-$r^2$* orbital symmetry.[10] In principle, no dichroic signal is expected for a cubic double exchange ferromagnetic manganite: the occurrence of a *$3z^2$-$r^2$* preferential occupation was already observed in very thin LSMO films and attributed to broken symmetry at the interface.[10]

On the base of the XAS results, a simple qualitative model is given in the following. Two different interfaces can be envisaged in these SLs: -CuO$_2$-Ca/MnO$_2$-La$_{0.7}$Sr$_{0.3}$O- and -Ca-CuO$_2$/La$_{0.7}$Sr$_{0.3}$O-MnO$_2$-. Ideally, the Ca- interface plane belongs to the IL structure. However, CaMnO$_3$ (perovskite structure) is also a stable compound so that the possibility can be envisaged of a -CuO$_2$/CaO-MnO$_2$-La$_{0.7}$Sr$_{0.3}$O- interface. Therefore, it is likely that the Ca- interface plane, under SOC, may accommodate a variable content $x$ of extra oxygen ions, $CuO_2$-$CaO_x$-$MnO_2$-$La_{0.7}Sr_{0.3}O$, which thus should provide doping holes to the SL and act as apical oxygens for Cu in the $CuO_2$ planes. This gives rise to Cu-O-Mn covalent bonding,[11] contributing to localize extra holes at the interface Cu $3d_{z2-r2}$ - Mn $3d_{z2-r2}$ hybrid orbitals. In agreement with the XAS spectra at the Cu $L_3$ edge, where no ligand hole peak is detected (Figs. 5 *a)* and *b)*), this interface orbital reconstruction hinders the hole doping of the $CuO_2$ layers and, thus, the occurrence of superconductivity. On the contrary, some extra hole doping of the LSMO block can be evinced from the appearance of the weak $Mn^{4+}$ peak in the *Mn $L_{2,3}$* absorption edge. Yunoki and coworkers[15] showed (Fig.3 of the cited work) that the top of the valence band in LSMO lies well above the top of the lower Hubbard band (valence band) of some cuprates such as $La_2CuO_4$, $Nd_2CuO_4$ and $Sm_2CuO_4$. In CCO, the gap is about 1.5 eV,[30] very similar to the other cuprates.[31] Since we expect the electronic structure of CCO to be not very different from the other mentioned cuprates, when extra holes, beyond those trapped at the interface, have to be located, they would likely occupy the LSMO valence band, rather than the CCO lower lying valence band. An increase of the hole concentration in the LSMO block beyond the optimal value ( $x \cong 0.3$ ) favours the stabilization of the C-type antiferromagnetic insulating phase,[32] thus increasing the thickness of the "dead layer". The opposite occurs in $CaCuO_2/SrTiO_3$ superlattices,[8] where the top of the valence band of $SrTiO_3$ lies well below the one of CCO.[15] In this case, indeed, the doping holes are transferred to the CCO block and superconductivity appears. Therefore, besides the chemical properties of the interfaces, which, as in the case of CCO/LSMO, can lead to orbital reconstruction and charge localization, also the band alignment between the two oxides in the heterostructure is a key factor to be taken into account. In particular, *wide gap semiconductors*, as $SrTiO_3$, are probably the most suitable systems to be coupled to CCO or to other insulating cuprates in order to obtain superconducting heterostructure.

In summary, we synthesized new high quality cuprate/manganite SLs using the infinite layer CCO as the cuprate block. The electrical transport properties of CCO/LSMO SLs coincide with those of LSMO bare films. In accordance with the results reported on ultrathin manganite films, the metal-insulator transition temperature is gradually reduced with decreasing the thickness of the LSMO block in the SLs, until a full insulating behaviour, corresponding to the stabilization of the C-type antiferromagnetic phase, is achieved. XAS measurements show that extra apical oxygen ions, forced at the interface between CCO and LSMO via a strongly oxidizing treatment, contribute to covalent bonding at the interface between *Cu $3d_{3z^2-r^2}$* and *Mn $3d_{3z^2-r^2}$* orbitals. This mechanism localizes holes at the interface preventing charge transfer to the CCO block. On the other hand, some charge transfer occurs toward the LSMO block contributing to the suppression of the magnetotransport properties which, in strongly oxidized SLs, occurs for larger values of the LSMO block thickness.

We are grateful to G. Ghiringhelli for very useful suggestions. This work was partly supported by the Italian MIUR (GrantNo.PRIN-20094W2LAY,"Ordine orbitale e di spin nelle eterostrutture di cuprati e manganiti").

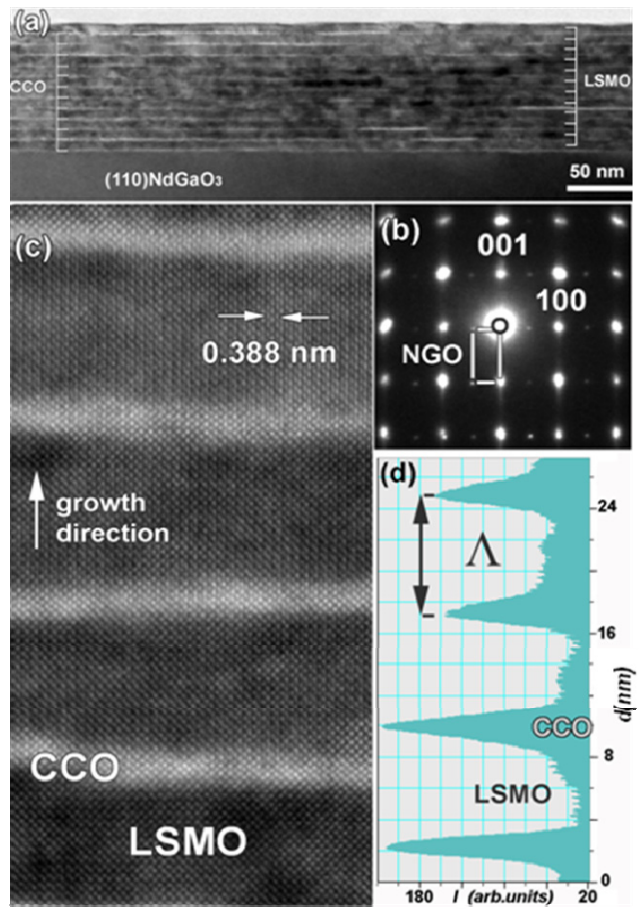

Fig.1 (Color online)(a) Cross-section bright field low magnification image of CCO/LSMO superlattice grown on (110)-NdGaO$_3$ substrate and (b) corresponding ED pattern. ED pattern is superposition of NdGaO$_3$ substrate (marked by white rectangle) and CCO - LSMO film indexed in pseudo-cubic structure; (c) selected area HRTEM image of CCO/LSMO superlattice and (d) corresponding intensity plot profile. The bright contrast layer and highest peak in plot profile (d) correspond to CCO layers.

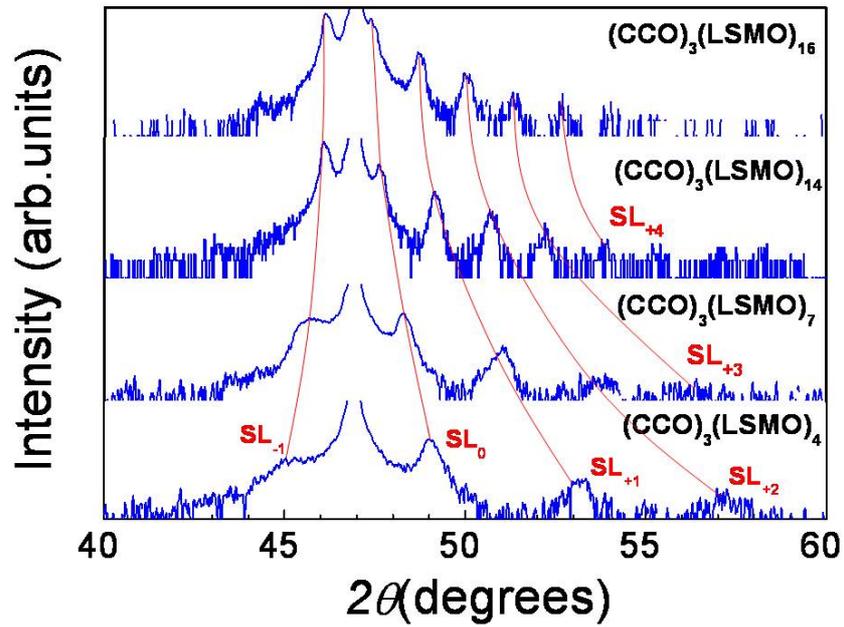

Fig.2 (Color online) X ray diffraction spectra around the (002) reflection peak of the NGO substrate of several $(CCO)_3/(LSMO)_n$ superlattices with different LSMO thickness $n$. The lines are guides for the eye and show how the superlattice peaks shift as the thickness of the LSMO block is varied.

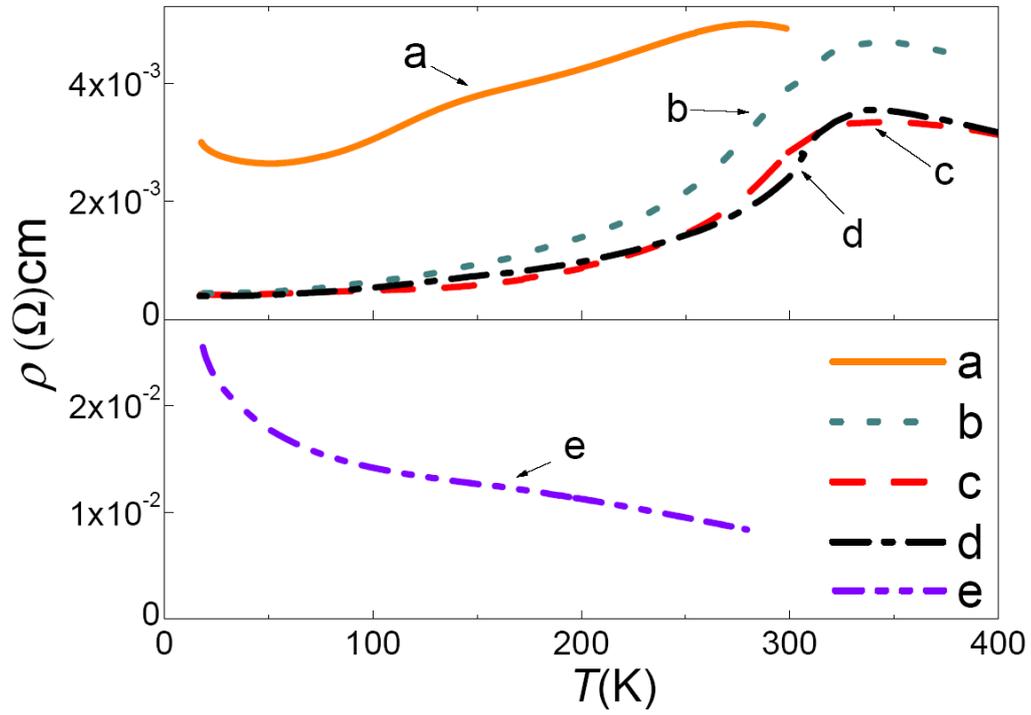

Fig.3 (Color online) Resistivity as a function of temperature for a) $CCO_3/LSMO_{14}$ superlattice grown under SOC, b) LSMO film, 15 unit cells thick, grown under SOC, c) $CCO_3/LSMO_{14}$ superlattice grown under MOC, d) LSMO film, 15 unit cells thick, grown under MOC, e) $CCO_9/LSMO_4$ superlattice grown under SOC.

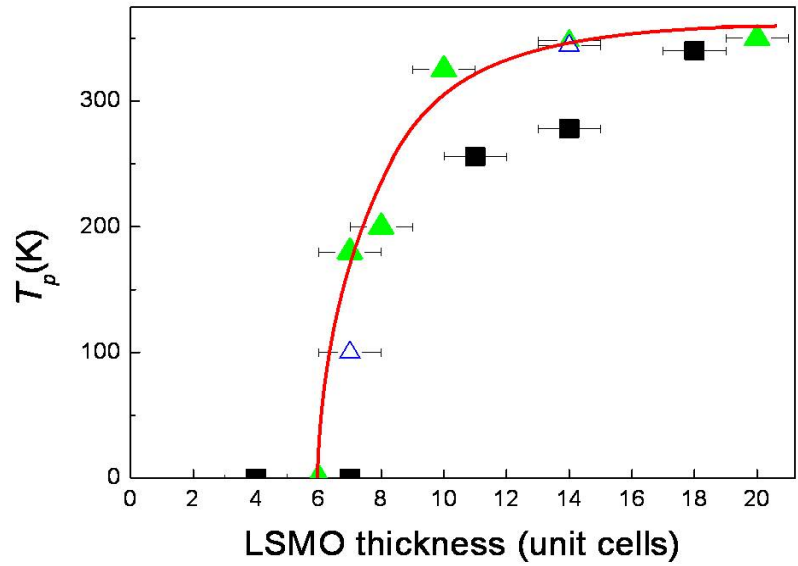

Fig.4 (Color online) Metal insulator transition temperature ($T_P$) as a function of the LSMO thickness for various LSMO films and $(CCO)_m/(LSMO)_n$ superlattices. The full line represents the behaviour of $T_p$ for bare LSMO films and has been drawn on the basis of data taken from ref.10 (not shown) and from this paper (full triangles). The full squares and empty triangles represent CCO/LSMO superlattices grown under SOC and MOC, respectively.

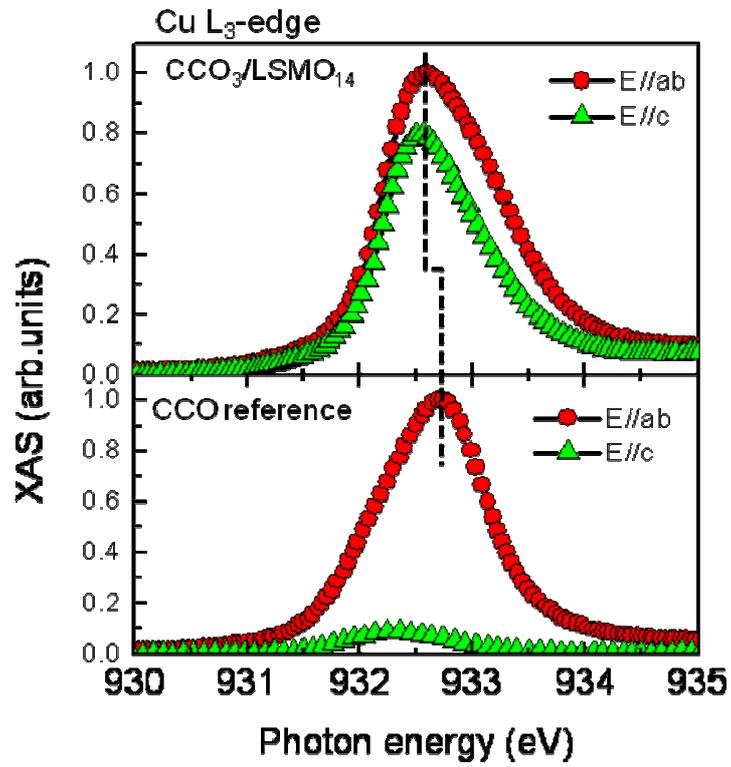

Fig.5 (Color online) Normalized XAS spectra taken at the Cu $L_3$-edge for a $CCO_3/LSMO_{14}$ superlattice (upper panel) and a reference CCO film (lower panel). Spectra are recorded both with the electric field perpendicular (triangles) and parallel (circles) to plane.

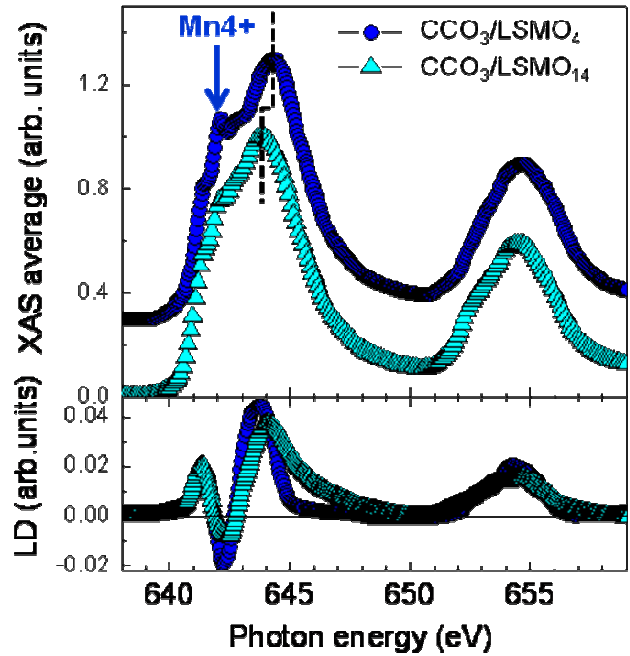

Fig.6 (Color online) Upper panel: absorption spectra at the Mn $L_{2,3}$ edges averaged between the signals taken with the electric field vector perpendicular and parallel to plane, for two SLs having different thickness of the LSMO block:14 u.c. (triangles) and 4 u.c. (circles). The average spectra are normalized to unity and vertically shifted for clarity. Lower panel: LD for the two SLs normalized to the average absorption.